\documentclass[aps,prl,twocolumn,superscriptaddress,showpacs]{revtex4}
\usepackage[ansinew]{inputenc}
\usepackage{epsfig}
\usepackage{graphicx}
\usepackage{color}
\usepackage{bm}
\usepackage{multirow}
\usepackage{mathrsfs}
\usepackage{revsymb}
\usepackage{amssymb}
\usepackage{amsmath}
\usepackage{slashed}
\usepackage{longtable}
\usepackage{dcolumn}
\newcommand{\kap}{\kappa}
\newcommand{\eps}{E}

\newcommand{\pot}{[\mbox{\boldmath $\alpha$} \times \mbox{\boldmath $r$}]_z }

\begin{document}
\preprint{Version 2.3.3}

\title{Nuclear Shape Effect on the $g$ Factor of Hydrogenlike Ions}

\author{Jacek Zatorski}
\email[]{jacek.zatorski@mpi-hd.mpg.de}
\affiliation{Max Planck Institute for Nuclear Physics, Saupfercheckweg~1, 69117 Heidelberg, Germany}

\author{Natalia S. Oreshkina}
\affiliation{Max Planck Institute for Nuclear Physics, Saupfercheckweg~1, 69117 Heidelberg, Germany}

\author{Christoph H. Keitel}
\affiliation{Max Planck Institute for Nuclear Physics, Saupfercheckweg~1, 69117 Heidelberg, Germany}

\author{Zolt\'an Harman}
\affiliation{Max Planck Institute for Nuclear Physics, Saupfercheckweg~1, 69117 Heidelberg, Germany}
\affiliation{ExtreMe Matter Institute EMMI, Planckstrasse 1, 64291 Darmstadt, Germany}

\date{\today}

\begin{abstract}

The nuclear shape correction to the $g$ factor of a bound electron in $1S$-state is calculated for a number of nuclei in the range of charge numbers from $Z=6$ up to $Z=92$. The leading relativistic deformation correction has been derived analytically and also
its influence on one-loop quantum electrodynamic terms has been evaluated. We show the leading corrections to become significant for mid-$Z$ ions and for very heavy elements to even reach the 10${}^{-6}$ level.

\end{abstract}

 \maketitle

The ever-increasing precision of measurements and theory of the $g$ factor of a bound electron has recently delivered
a new value for the electron mass~\cite{Haffner2000,Verdu2004}, and keeps providing stringent tests for quantum electrodynamics (QED)
in strong fields~\cite{Haffner2000,Verdu2004,Sturm2011}. It also allows to access electromagnetic properties of nuclei such as
charge radii, as demonstrated in a very recent proof-of-the-principle study with a
$\textrm{Si}^{13+}$ ion~\cite{Sturm2011}, or, as suggested theoretically, magnetic moments~\cite{Yerokhin2011}. Also, it is anticipated that $g$ factor studies will yield a value for the fine-structure constant $\alpha$ that is more accurate than the presently established one
when extending the experiments to elements with a high charge number $Z$~\cite{Shabaev2006}. 

In a few years, measurements with the heaviest elements will be possible~\cite{HITRAP2008}.
As higher-order nuclear and QED contributions to the theoretical value of the $g$ factor are strongly boosted with increasing $Z$, at the present
10${}^{-10}$ level of relative experimental accuracy~\cite{Sturm2011, Sturm-Phase2011} or even below, our present understanding of atomic structure
will not be satisfactory. In such strong Coulomb fields, nuclear effects beyond a simple spherical model arise. Furthermore, QED and nuclear structural
contributions are intertwined. 

In this Letter we consider the nuclear shape effect, and find that while it can be safely neglected in
predictions for low-$Z$ systems, it greatly influences the $g$ factor value already for mid-$Z$
elements. At high nuclear charges, its inclusion in the theoretical description is mandatory; for example, for U${}^{91+}$, its relative
contribution to the total $g$ factor reaches the 10${}^{-6}$ level. We furthermore evaluate mixed nuclear-QED terms, i.e. the nuclear shape effect
on the one-loop QED terms of self-energy (SE) and vacuum polarization (VP). Even these contributions will be highly relevant for the interpretation
of experimental values to be obtained within a few years~\cite{HITRAP2008}. Furthermore, a comparison of theory and experiment may even yield more accurate values for nuclear shape parameters, relevant in explaining shape phase transitions in nuclear structure theory~\cite{Cejnar2010}.

We account for the nuclear quadrupole and hexadecapole deformation and
derive a formula describing the nuclear shape correction to the $g$ factor of a bound electron
in hydrogen-like ions in the $1S$-state. Then we focus on systems where the nucleus is spinless and in its ground state.
Let us start with the definition
of the electron $g$ factor (we use units with $c=1$, $\hbar=1$, $\alpha = e^2/(4\pi)$, and with $e=-\left|\,e\right| \,$ unless otherwise stated)
\begin{equation}
\label{def:gfactor}
	\delta E = -\frac{e}{m} \, \bigl< \vec{s} \vec{B} \bigr> \, \frac{g}{2} \, ,  
\end{equation}
where $\delta E$ stands for the energy correction due to the coupling of the electron spin operator $\vec{s}$ to the external magnetic field $\vec{B}$, $m$ is the mass of the electron, and $ \bigl< \ldots \bigr>$ stands for the expectation value.
The nuclear deformation (ND) correction to the $g$ factor, which we present, is closely related to the finite-size (FS) effect, which is well known in the context of atomic levels. The correction emerges when one accounts for a deviation of a nucleus from the spherical shape. 

Let us consider now a relativistic Hamiltonian of the bound system of an electron and a nucleus in the presence of an external, constant and homogeneous magnetic field. The interaction of the nucleus with the external magnetic field is negligible so that we can write
\begin{equation}
	H = H_a + H_N + \delta V_{C} - e \vec \alpha \vec{A}  \, ,
\end{equation}
where $	\vec{A} = \frac{1}{2}\,(\vec{B} \times \vec{r})$, $H_a$ is the atomic Dirac-Coulomb Hamiltonian $H_a = \vec \alpha \vec{p} + \beta m - \frac{Z \alpha}{r}$, $\vec{r}$ is the vector describing the electron's position with respect to the mass-center of the system, $H_N$ the (unknown) nuclear Hamiltonian and $\delta V_{C}$ the deviation from the monopole term expansion of the exact Coulomb potential of our system, i.e.
\begin{equation}
\label{deltaV}
	\delta V_{C} = \frac{Z \alpha}{r} - \sum_{i=1}^{Z} \frac{\alpha}{|\vec{r} - \vec{r}_i|} \, .
\end{equation}  
Now we formulate the framework for the perturbative calculation of the energy correction due to the presence of the magnetic field and
the deviation from the Coulomb potential. We start with the equation
\begin{equation}
\label{eq:eigenenergy}
	\bigl( H_a + H_N + \delta V_{C} - e \vec \alpha \vec{A} \bigr) \left| \Psi \right> = (E_{a}^{(0)} + E_{N}^{(0)} + \delta E) \left| \Psi \right> \, ,
\end{equation}
where $\left| \Psi \right>$ denotes the perturbed state of the system, $E_{a}^{(0)}$ the unperturbed ground-state energy of the atom, $E_{N}^{(0)}$ the unperturbed ground-state energy of the nucleus and $\delta E$ denotes the energy correction arising due to $\delta V_{C}$ and the interaction with the external magnetic field. Next, we perform the following approximation: we assume that the potential $\delta V_C$ does not induce any nuclear transition, which means that we neglect the nuclear polarizability (NP) effects. However, the derivation of the NP correction as well as its numerical values for a number of elements can be found in Ref.~\cite{Nefiodov}. In principle there are also mixed terms containing both: the NP and nuclear deformation effects, but since these effects are both already small we neglect the mixed terms.
Then we can write
$\left| \Psi \right> = \left| \phi' \right> \otimes \left| \psi \right>$,
where $\left| \psi \right>$ stands for the ground state of the nucleus, $\left| \phi' \right>$ denotes the perturbed atomic state
$	\left| \phi' \right> = \left| \phi \right> + \left| \delta \phi \right>$
and $\left| \delta \phi \right>$ stands for the correction to the unperturbed atomic state $\left| \phi \right>$. We substitute the state $\left| \Psi \right>$ into Eq.~(\ref{eq:eigenenergy}) and act on the resulting equation with $\big< \psi \big|$. Following an obvious reduction we obtain
\begin{equation}
	\label{eq:eigenenergy2}
	\bigl( H_a + \delta \hat{V}(\vec{r}) - e \vec \alpha \vec{A} \bigr) \left| \phi' \right> = (E_{a}^{(0)} + \delta E) \left| \phi' \right> \, ,
\end{equation}
where $\delta \hat{V}(\vec{r}) \equiv \left< \psi \right| \delta V_{C} \left| \psi \right>$.

Now we will express $\delta \hat{V}(\vec{r})$ in terms of a nuclear charge distribution $\rho(\vec{r})$. We do not focus on details here since the detailed derivation of the formula~(\ref{eq:integralV3}) can be found in Ref.~\cite{Kozhedub}. If the nucleus is in the state of vanishing angular momentum $I=0$, then the potential $\delta \hat{V}(\vec{r})$ can be written in terms of the simple formula 
\begin{equation}
\label{eq:integralV3}
	\delta \hat{V}(r) = \frac{Z\alpha}{r} - e k  \int dr' \, {r'}^{2} \, \frac{\rho(r')}{r_{>}} \, ,
\end{equation}
where $r_{>} = \max(r, r')$. In Eq.~(\ref{eq:integralV3}) we have introduced the radial charge density, i.e.
$\rho(r) = \int d\vec{n} \, \rho(\vec{r})$.
The terms on the right-hand side (RHS) of Eq.~(\ref{eq:integralV3}) corresponds to the respective terms on the RHS of Eq.~(\ref{deltaV}). 
Let us note that the resulting potential $\delta \hat{V}(r)$ depends solely on the radial variable $r$. 

In our calculation we consider axially symmetric nuclei and employ the two-parameter Fermi charge distribution with quadrupole and hexadecapole deformation, i.e.
\begin{equation}
\label{def:rho}
	\rho(\vec r) = \frac{N}{1 + \exp{\left[(r - c)/a \right]}} \, ,
\end{equation}
where $N$ is a normalization constant, the half--density radius is given in terms of spherical harmonics as
$c = c_{0} (1 + \beta_2 Y_{20}+ \beta_4 Y_{40})$ with $\beta_{2}$ and $\beta_{4}$ being the octupole and hexadecapole deformation parameters, respectively.

The $g$ factor corrections of interest to us for the systems described by Eq.~(\ref{eq:eigenenergy2}) might be calculated with the help of the results of Ref.~\cite{Karshenboim2005}. We will use Eq.~(27) of Ref.~\cite{Karshenboim2005}, which for a $1S$-state reads
\begin{equation}
\label{eq:gfns}
	\delta g_{\textrm{FS}} = \frac{4 (2\gamma +1)}{3} \, \frac{E_{\textrm{FS}}}{m} \, .
\end{equation}
This formula describes a correction to the $g$ factor due to the FS effect for arbitrary radial distributions of the nuclear charge.
$E_{\textrm{FS}}$ denotes the energy shift due to the FS effect described by the perturbing scalar potential, and
$\gamma = \sqrt{1 - (Z\alpha)^2}$. Eq.~(\ref{eq:gfns}) takes into account the interaction with the external magnetic field in the first order of perturbation theory, whereas the interaction with $\delta \hat{V}(r)$ is treated to all orders. In practice, the accuracy of Eq.~(\ref{eq:gfns}) is limited by the accuracy of $E_{\textrm{FS}}$.  
In order to obtain $E_{\textrm{FS}}$ we use Eq.~(17) of Ref.~\cite{Shabaev1993}. Namely, for the $1S$-state
\begin{equation}
\label{eq:Efns}
	E_{\textrm{FS}} = \frac{(\alpha Z)^2}{10} \, \bigl[ 1 + (\alpha Z)^2 f(\alpha Z)\bigr] \, \biggl(2 \alpha Z R m \biggr)^{2 \gamma} \, m \, ,
\end{equation}
where $f(\alpha Z) = 1.380 - 0.162 (\alpha Z) + 1.162 (\alpha Z)^2$, 
and $R$ is the effective radius of the homogeneously charged sphere that gives the same energy correction as the original nuclear shape. For states with $j=1/2$ 
\begin{equation}
\label{eq:Reffective}
	R = \sqrt{\frac{5}{3} \bigl< r^2 \bigr> \biggl[ 1 - \frac{3}{4} (\alpha Z)^2 \biggl(\frac{3}{25}\bigl< r^4\bigr>/\bigl< r^2\bigr>^2 - \frac{1}{7} \biggr) \biggr] } \, ,
\end{equation}
where $	\bigl< r^n \bigr> \equiv {\int dr \, r^2 \, r^n \rho(r)}/{(Z |e|)}$. 
In our calculation $\bigl< r^4 \bigr>$ is computed by numerical integration, whereas $\bigl< r^2 \bigr>$ is found in the literature~\cite{Angeli}. 

It might be instructive to obtain the approximate formula expressing $R$ in terms of the nuclear parameters. For this purpose, we substitute relations (24)-(26) from Ref.~\cite{Kozhedub} into Eq.~(\ref{eq:Reffective}) and expand the result. In this way we obtain
\begin{eqnarray}
\label{Effective-R}
	R &\simeq& \bigl< r^2 \bigr>^{1/2} \biggl\{ \sqrt{\frac{5}{3}} -  \frac{(Z\alpha)^2}{14 \pi} \biggl[ \frac{\sqrt{15}}{4} \, \beta_{2}^{2}  + \frac{5 \sqrt{3}}{7 \pi^{1/2}} \, \beta_{2}^{3} + \nonumber   \\ 
	&&   \frac{9 \sqrt{15}}{7 \pi^{1/2}} \, \beta_{2}^{2} \beta_4  + \left(\frac{a}{c}\right)^2 \biggl( \sqrt{\frac{5}{3}} \, \pi^2  - \frac{147 \sqrt{5} \pi^2}{28 \sqrt{3}} \, \beta_{2}^2 -  \nonumber \\
 	&&  \frac{355 \pi^{3/2}}{28 \sqrt{3}} \, \beta_{2}^3 - \frac{639 \sqrt{5} \pi^{3/2}}{28 \sqrt{3}} \, \beta_{2}^2 \beta_4 
 	\biggr) \biggr] \biggr\} . 
\end{eqnarray} 
The values of $R$ obtained with the help of numerical integration are in good agreement with those obtained with formula~(\ref{Effective-R}).
Finally, the nuclear deformation correction to the $g$ factor is defined as
\begin{equation}
\label{def:deltagND}
	\delta g_{\textrm{ND}} \equiv \delta g_{\textrm{FS}}\bigl(Z, \sqrt{\bigl< r^2 \bigr>}, a, \beta_{2}, \beta_{4}\bigr) - \delta g_{\textrm{FS}}\bigl(Z, \sqrt{\bigl< r^2 \bigr>}, a, 0, 0 \bigr) \, ,
\end{equation}
where $\delta g_{\textrm{FS}}$ depends on the nuclear parameters via $R$.

We also take into account nuclear deformation correction to the QED corrections. 
The QED corrections to the $g$ factor of first order in $\alpha$ consist of self-energy (SE) and vacuum-polarization (VP) corrections.
We mostly use the calculation scheme as in~\cite{glazov_pla_2006,oresh_pla_2007}. The 
SE term represents the sum of irreducible, reducible and vertex parts:
$\delta g_{\rm{SE}} = \delta g_{\rm{irr}} + \delta g_{\rm{red}} + \delta g_{\rm{ver}}$.
The irreducible part is given by~\cite{shabaev-pr}
\begin{align} \label{irr}
\delta g_{\rm{irr}} = \frac{1}{m_\phi} \sum_{n}^{\eps_n \neq \eps_\phi} \frac{\langle \phi| (\Sigma(\eps_\phi) - \gamma^0\delta m)
	  |n\rangle \langle n | \pot | \phi \rangle} {\eps_\phi - \eps_n}.
\end{align}
Here, $m_\phi$  is  an angular momentum of the state $|\phi\rangle$ and $\eps_n$ 
the energy of state $|n\rangle$, $\pot$ describes the
interaction with the external magnetic field, and $\delta m$ is a mass counterterm. $\Sigma(\eps)$
denotes the unrenormalized self-energy operator defined as
\begin{align}
\langle a| \Sigma(\eps)|b\rangle   = \frac{i}{2\pi} \int d\omega \sum_n 
	\frac {\langle an |I(\omega)|nb\rangle}{\eps - \omega - \eps_n(1-i0)},
\end{align}
where $I(\omega,x_1,x_2) = e^2 \alpha^\mu \alpha^\nu D_{\mu\nu}(\omega,x_1,x_2)$ with the Dirac matrices 
$\alpha^\mu = (1,\alpha)$, and the photon propagator $D_{\mu\nu}(\omega,x_1,x_2)$. 
The expressions for the reducible and vertex parts read \cite{shabaev-pr}
\begin{align} \label{vr}
 \delta g_{\rm{red}} &=  \frac{1}{m_\phi} \langle \phi | \pot | \phi \rangle 
	  \langle \phi| \frac{d}{d\eps}\Sigma(\eps)|_{\eps = \eps_\phi} |\phi\rangle\,, \\
 \delta g_{\rm{ver}} &= \frac{1}{m_\phi} \frac{i}{2\pi} \int d\omega \sum_{n_1,n_2} \\
	& \frac {\langle \phi n_2 |I(\omega)|n_1\phi\rangle \langle n_1 | \pot | n_2 \rangle}
	{(\eps_\phi - \omega - \eps_{n_1}(1-i0))(\eps_\phi - \omega - \eps_{n_2}(1-i0))}\,. \nonumber 
\end{align}
Both the reducible and vertex parts are ultraviolet-divergent, whereas the sum 
$\delta g_{\rm{vr}} = \delta g_{\rm{red}} + \delta g_{\rm{ver}}$
is finite. To separate the ultraviolet divergencies, the expression \eqref{irr} is decomposed into zero-, one-, and many-potential
terms,
and the expression \eqref{vr} is decomposed into zero- and many-potential terms. All zero- and one-potential terms are evaluated 
in momentum space. The remaining many-potential parts $\delta g_{\rm{irr}}^{(2+)}$ and $\delta g_{\rm{vr}}^{(1+)}$ are calculated 
in coordinate space \cite{blundell}. 
The angular integration and the summation over intermediate angular
projections in the many-potential terms $\delta g_{\rm{irr}}^{(2+)}$ and $\delta g_{\rm{vr}}^{(1+)}$ are 
carried out in a standard algebraic manner. The many-potential terms involve infinite summation over the
quantum  numbers $\kap = \pm(j+1/2)$. This summation is terminated at a maximum value
$|\kap| = 15-20$, and the residual part of the sum is evaluated by a least-square inverse-polynomial fitting.
The summation over the Dirac spectrum at fixed intermediate $\kap$ was carried out with the dual-kinetic-basis (DKB) method~\cite{dkb}.

The VP correction consists of two
parts, originating from the vacuum-polarization diagram with the magnetic 
interaction inserted into the external electron line (the electric-loop contribution $\delta g_{\rm{VP}}^e$) 
and into the vacuum-polarization loop (the
magnetic-loop contribution $\delta g_{\rm{VP}}^m$). We took into account only the  
leading term since it guarantees the accuracy needed in the present work. The leading term originates from the electric loop and can be described by the Uehling potential. The VP correction is calculated using the DKB method as well.

To calculate the nuclear deformation correction to QED corrections (QED-ND), we used an effective radius of the nucleus.
We employed the shell, homogeneously charged sphere and Fermi nuclear distributions. 
Although the QED-ND-calculation results are numerically stable and do not depend of the nuclear model, they should be considered 
as an estimate.

\begin{table}[!ht]
\caption{The nuclear deformation correction to the $g$ factor ($\delta g_{\textrm{ND}}$). The values of $\bigl< r^2 \bigr>^{1/2}$ and their uncertainties originate from Ref.~\cite{Angeli}. Unless otherwise stated the value of $a$ was obtained by equation $a = t/(4 \log{3})$ (Ref.~\cite{Beier2000}), where $t = 2.30$ fm and rather conservative error bars were assumed.
The parameters $\beta_2$ were estimated with the help of formula~(\ref{beta2}) unless a reference is given. For $\null^{234}\textrm{U}$ and $\null^{238}\textrm{U}$ we employed, in accordance with~\cite{Zumbro}, $\beta_4 = 0.08(5)$ and $\beta_4 = 0.07(5)$, respectively; for Si, in accordance with~\cite{Rebel}, $\beta_4 = 0.08(5)$; for other elements we assumed $\beta_4 = 0.0(1)$. The values of $c_0$ are set by the requirement that the numerically computed $\bigl< r^2 \bigr>^{1/2}$ should be equal to the established ones~\cite{Angeli}.
See also the comments in the text on the uncertainty of $\delta g_{\textrm{ND}}$.}
	\label{tab:table1}
		\begin{ruledtabular}
		\begin{tabular}{llllll}
  Z & Isotope   &  $a$ (fm)& $\beta_2$ & $\delta g_{\textrm{ND}}$   \\ \hline
  6 & $\null^{12} \textrm{C}$  & 0.523(40) 	& \phantom{-}0.44(10) 	    &  $ -7.9(5.3)  \cdot 10^{-16}$   \\
 14 & $\null^{28} \textrm{Si}$ & 0.523(20) 	& -0.349(20)\footnotemark[3] &  $ -2.85(52)  \cdot 10^{-13}$   \\
 	  & $\null^{30} \textrm{Si}$ & 0.523(20)   & -0.314(20)\footnotemark[3]&  $ -2.48(49)  \cdot 10^{-13}$   \\
 38 & $\null^{86} \textrm{Sr}$ & 0.523(20)   & \phantom{-}0.134(10) &  $ -9.0(3.1)   \cdot 10^{-11}$   \\
    & $\null^{100}\textrm{Sr}$ & 0.523(20)   & \phantom{-}0.435(11) &  $ -1.08(28)  \cdot 10^{-9} $   \\
 60 & $\null^{142}\textrm{Nd}$ & 0.523(20)	  & \phantom{-}0.100(20) &  $ -2.0(1.1)  \cdot 10^{-9}$    \\
    & $\null^{150}\textrm{Nd}$ & 0.523(20)	  & \phantom{-}0.278(20)\footnotemark[1]  &  $ -1.70(53)  \cdot 10^{-8}$    \\
 62 & $\null^{144}\textrm{Sm}$ & 0.523(20)	  & \phantom{-}0.090(20)\footnotemark[1]  &  $ -2.1(1.2)   \cdot 10^{-9}$    \\
    & $\null^{154}\textrm{Sm}$ & 0.498(20)\footnotemark[1]	  & \phantom{-}0.328(20)\footnotemark[1]   &  $ -3.24(98)  \cdot 10^{-8}$    \\
 92 & $\null^{234}\textrm{U} $ & 0.509(20)\footnotemark[2]	  & \phantom{-}0.256(10)	&  $ -1.12(27)  \cdot 10^{-6}$    \\
    & $\null^{238}\textrm{U} $ & 0.505(20)\footnotemark[2]   & \phantom{-}0.280(10)  &  $ -1.28(28)  \cdot 10^{-6}$    \\ 
		\end{tabular}
		\end{ruledtabular}
\footnotetext[1]{Ref.~\cite{Fricke}.}
\footnotetext[2]{Ref.~\cite{Zumbro}.}
\footnotetext[3]{The sign originates from Ref.~\cite{Rebel}.}
\end{table}

Table~(\ref{tab:table1}) presents numerical values of $\delta g_{\textrm{ND}}$ obtained with formula~(\ref{def:deltagND}) for various hydrogenlike ions. Depending on the availability of experimental data, the quadrupole deformation parameter $\beta_{2}$ was either taken from literature or estimated with the help of the formula~\cite{Trager}
\begin{equation}
\label{beta2}
	\beta_{2} = \frac{4 \pi}{3 Z |e| R_{s}^2} \, \biggl[ \sum_{i} B(E2; 0^{+} \rightarrow 2_{i}^{+}) \biggr]^{1/2} \, ,
\end{equation}
where: $B(E2; 0^{+} \rightarrow 2_{i}^{+})$ stands for the reduced probability of $E2$ transition from the ground state $0^{+}$ to a state $2_{i}^{+}$,
$R_{s}$ is the effective radius evaluated with the help of the simplified formula 
$R_{s} = \sqrt{\frac{5}{3} \, \bigl< r^2 \bigr>}$, and the values of $B(E2; 0^{+} \rightarrow 2_{i}^{+})$ were evaluated with the help of the data from Ref.~\cite{ENSDF}. The uncertainty of $\delta g_{\textrm{ND}}$ is estimated as the quadratic sum of uncertainties related to the nuclear parameters.
We assume rather conservative error bars for the parameters $a, \beta_{2}$ and $\beta_4$ to account for an uncertainty caused by the fact that for some nuclei the parameters were compiled from various experimental results.

As for the QED-ND effect, it is significantly smaller than the leading ND correction. For example, we obtained the following values: $5(3) \cdot 10^{-9}$, $4(2) \cdot 10^{-9}$, $1.2(4) \cdot 10^{-10}$, $1.3(1.1) \cdot 10^{-11}$, $1.5(3) \cdot 10^{-11}$, and $1.6(6) \cdot 10^{-12}$ for $\null^{238}\textrm{U}$, $\null^{234}\textrm{U}$, $\null^{150}\textrm{Nd}$, $\null^{142}\textrm{Nd}$, $\null^{100}\textrm{Sr}$, and $\null^{86}\textrm{Sr}$, respectively. The accuracy of the results for the ND and QED-ND corrections is such that the comparison with experimental values of the $g$ factor could serve as a method for the determination of $\beta_2$ and $\beta_4$ for heavy nuclei, provided that all other relevant corrections as well as remaining nuclear parameters are known with sufficient accuracy.

Table~(\ref{tab:table2}) contains the values of the isotopic shift of $\delta g_{\textrm{ND}}$ for various chosen isotopes.
For a comparison with experimental results, values of other isotope-dependent effects are required.
The leading FS effect as well as the mixed FS-VP effect can be found in~\cite{Beier2000}, the mixed FS-SE correction in~\cite{Yerokhin2004}, the recoil effects in~\cite{Shabaev2002, Eides1997, Grotch1971}, and the NP effect in~\cite{Nefiodov}.

\begin{table}[!ht]
\caption{The isotopic shifts of various corrections to the $g$ factor.
The following notation for an isotopic shift has been introduced: $\Delta g_{\textrm{X}} \equiv \delta g_{\textrm{X}}(A_1) - \delta g_{\textrm{X}}(A_2)$, where $A_1$ and $A_2$ are the first and second mass numbers, respectively, given in the second column, and $\textrm{X} \in \{\textrm{ND}, \textrm{ND-QED}\}$.}
	\label{tab:table2}
		\begin{ruledtabular}
		\begin{tabular}{lllll}
  Z & Isotopes  & $\Delta g_{\textrm{ND}}$ & $\Delta g_{\textrm{ND-QED}}$ & $\Delta g_{\textrm{ND, Total}}$  \\ \hline
 38 & $\null^{100, 86}\textrm{Sr} $ & $-9.9(2.8) \cdot 10^{-10}$ & $1.4(3) \cdot 10^{-11}$ & $-9.8(2.8)\cdot 10^{-10}$ \\
 60 & $\null^{150, 142}\textrm{Nd}$ & $-1.50(54) \cdot 10^{-8}$ & $1.1(4)\cdot 10^{-10}$ & $-1.49(54)\cdot 10^{-8}$ \\
 92 & $\null^{238, 234}\textrm{U}$ & $-1.6(3.9) \cdot 10^{-7}$	  & $1(4) \cdot 10^{-9}$ & $-1.6(3.9)\cdot 10^{-7} $\\
		\end{tabular}
		\end{ruledtabular}
\end{table}

\begin{figure}
	\centering
		\includegraphics[width=0.5\textwidth]{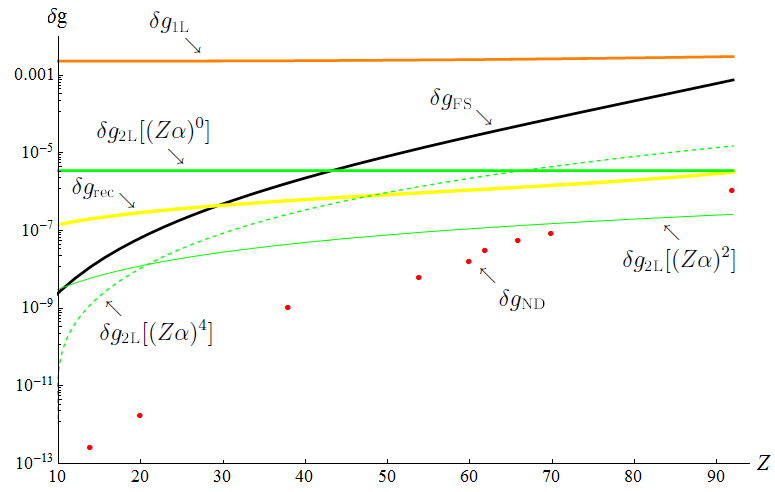}
	\caption{Various contributions to the $g$ factor: $\delta g_{\textrm{1L}}$ stands for one-loop all-order QED corrections, 
	$\delta g_{\textrm{FS}}$ for the finite-size effect, $\delta g_{\textrm{rec}}$ for the recoil correction,
	$\delta g_{\textrm{ND}}$ for the nuclear deformation correction and $\delta g_{\textrm{2L}}[(Z \alpha)^n]$ represents two-loop 
	QED corrections at order $(Z \alpha)^n$ (see~\cite{Pachucki2005}). }
	\label{fig:Plot_article}
\end{figure}

For comparison, Fig.~\ref{fig:Plot_article} presents a plot of $\delta g_{\textrm{ND}}$ and some other contributions to the $g$ factor. It is apparent that $\delta g_{\textrm{ND}}$ grows quickly with $Z$ and that it becomes prominent for heavy ions. 

In summary, we have developed the theory of the nuclear deformation correction to the bound electron $g$ factor and calculated this term for a wide range of elements. It turns out that already for $\null^{100} \textrm{Sr}^{37+}$, the relative nuclear shape contribution exceeds $10^{-10}$, which is the level that has just been reached in experiment~\cite{Sturm2011}.
For high-$Z$ elements, the correction becomes large, e.g. on the level of $10^{-6}$ for $\null^{238} \textrm{U}^{91+}$, and definitely will be important for the comparison between theory and experiment for high-$Z$ elements, which are expected within a few years~\cite{HITRAP2008}. These results are likely to allow in future for the extraction of nuclear deformation parameters from experimental values of the $g$ factor.

JZ and ZH acknowledge helpful conversation with A.~I. Milstein. The work of ZH has been supported by the Alliance Program of the Helmholtz Association (HA216/EMMI).

\end{document}